\documentclass[aps,pra,onecolumn,notitlepage,superscriptaddress]{revtex4-1}

\usepackage{xcolor,graphicx}

\begin{document}

\title{Fast Escape from Quantum Mazes in Integrated Photonics}

\author{Filippo Caruso}
\affiliation{LENS \& Dipartimento di Fisica e Astronomia, Universit\'{a} di Firenze, I-50019 Sesto Fiorentino, Italy}
\affiliation{QSTAR, Largo Enrico Fermi 2, I-50125 Firenze, Italy}
\affiliation{Istituto Nazionale di Ottica, Consiglio Nazionale delle Ricerche (INO-CNR), Largo Enrico Fermi 6, I-50125 Firenze, Italy}
\author{Andrea Crespi}
\affiliation{Istituto di Fotonica e Nanotecnologie, Consiglio Nazionale delle Ricerche
(IFN-CNR), Piazza Leonardo da Vinci 32, I-20133 Milano, Italy}
\affiliation{Dipartimento di Fisica, Politecnico di Milano, Piazza Leonardo da Vinci 32, I-20133 Milano, Italy}
\author{Anna Gabriella Ciriolo}
\affiliation{Dipartimento di Fisica, Politecnico di Milano, Piazza Leonardo da Vinci 32, I-20133 Milano, Italy}
\author{Fabio Sciarrino}
\affiliation{Dipartimento di Fisica, Sapienza Universit\'{a} di Roma, Piazzale Aldo Moro 5, I-00185 Roma, Italy}
\author{Roberto Osellame}
\affiliation{Istituto di Fotonica e Nanotecnologie, Consiglio Nazionale delle Ricerche
(IFN-CNR), Piazza Leonardo da Vinci 32, I-20133 Milano, Italy}
\affiliation{Dipartimento di Fisica, Politecnico di Milano, Piazza Leonardo da Vinci 32, I-20133 Milano, Italy}

\begin{abstract}
Escaping from a complex maze, by exploring different paths with several decision-making branches in order to reach the exit, has always been a very challenging and fascinating task. Wave field and quantum objects may explore a complex structure in parallel by interference effects, but without necessarily leading to more efficient transport. Here, inspired by recent observations in biological energy transport phenomena, we demonstrate how a quantum walker can efficiently reach the output of a maze by partially suppressing the presence of interference. In particular, we show theoretically an unprecedented improvement in transport efficiency for increasing maze size with respect to purely quantum and classical approaches. In addition, we investigate experimentally these hybrid transport phenomena, by mapping the maze problem in an integrated waveguide array, probed by coherent light, hence successfully testing our theoretical results. These achievements may lead towards future bio-inspired photonics technologies for more efficient transport and computation.
\end{abstract}

\maketitle

\begin{figure}[b]
\centerline{\includegraphics[width=.6\textwidth]{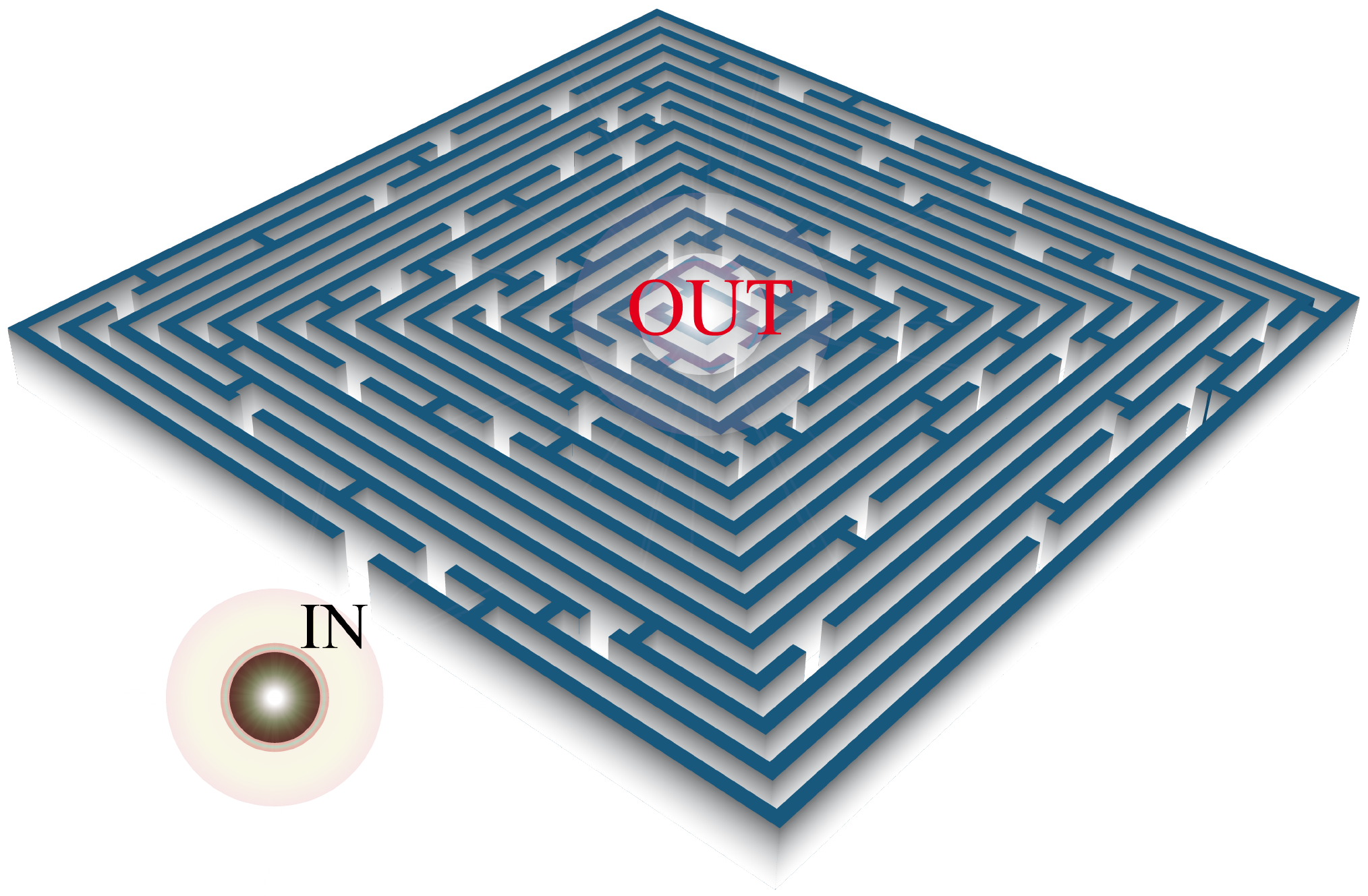}}
\caption{Pictorial view of a maze with single input (IN) and output (OUT) ports. An ideal walker has to travel from IN to OUT in the shortest possible time.}\label{fig1}
\end{figure}

Transport problems are very popular in several fields of science, as biology, chemistry, sociology, information science, physics, and even in everyday life. One of the most challenging transport problems is represented by efficiently traversing a maze, i.e. finding the exit in the shortest possible time of a topologically-complex network of interconnected sites (Fig. \ref{fig1}). The efficiency in reaching the exit of a maze dramatically decreases with the number of sites in the structure, rapidly making this problem intractable \cite{rivest76}.

The problem of solving mazes has fascinated mankind since the ancient times. One famous maze is the Cretan one, designed by the architect Daedalus, build to hold the mythological creature Minotaur that was eventually killed by the hero Theseus. To find the Minotaur he used the most typical maze-solving strategy: exploring several possible alternatives, while marking the visited paths (by a ball of thread). Around 60 years ago, Shannon realized the first ever experiment on maze-solving that was based on physical means, in particular an electromagnetic mouse Theseus \cite{shannon}. Nowadays, the availability of new physical, chemical and biological systems has opened up the way for traversing a maze with a parallel exploration of all possible transport channels at the same time. For instance, in Ref. \cite{STS95} a maze is experimentally solved by filling it with a Belousov-Zhabotinsky reaction mixture and then exploiting the superposition effect of travelling chemical wavefronts. More recently, it was shown that this parallel addressing can be indirectly obtained by the chemo-attractants emitted by the oat flake placed at the destination site, while a plasmodium slime walks directly to the exit \cite{Adamatzky}. This demonstrates the crucial role of interference to find the maze's exit in a more efficient way.

In the framework of quantum mechanics, even a single particle, represented by a wavefunction, shows interference effects. Exploiting this property, a quantum walker is able to propagate in the fastest way inside perfectly ordered lattices \cite{aharonov,Kempe06}; however, localization phenomena may occur when disorder is present \cite{KLMW2007,schwartz2007,lahini2008}. Quantum walks find applications to energy transport \cite{blumen} and quantum information \cite{santha,farhi,childs,childs09} with polynomial as well as exponential speed-up \cite{ambainis}, e.g. Grover search algorithm \cite{grover}, universal models for quantum computation \cite{childs13}, state transfer in spin and harmonic networks \cite{bose,ekert,PHE04}, and recent proposals on web page ranking \cite{zueco12}. Recently, the maze problem has been converted into a quantum search problem to get a quadratic speedup \cite{KG2013}. Interestingly enough, the interplay of interference and noise effects can further enhance quantum transport over complex networks, as recently observed for energy transport phenomena in light-harvesting proteins \cite{engel07,MRLA2008,PH2008,CCDHP2009,caruso14} and proposed for noise-assisted quantum communication \cite{CHP2010}.
In the last years, several technological platforms have been employed to investigate quantum transport phenomena, such as NMR \cite{du,ryan}, trapped ions \cite{schmitz,zahringer}, neutral atoms \cite{karski}, and several photonic schemes as bulk optics \cite{broome,white}, fiber loop configurations \cite{schreiber12,jeong}, and miniaturized integrated waveguide circuits \cite{perets,peruzzo,owens,sansoni,crespi}. Among these, a very interesting experimental platform is represented by three-dimensional waveguide arrays, fabricated by femtosecond laser micromachining \cite{owens,CresNJP,SzamNat,CorrNC}. Femtosecond laser waveguide writing \cite{dellavalle2009} enables to fabricate high-quality optical waveguides, directly buried in the bulk of a transparent substrate. Ultrashort laser pulses are focused at the desired depth in the substrate and nonlinear absorption processes induce localized and permanent refractive index increase; translation of the sample at uniform speed allows to draw guiding paths in the substrate with unique three-dimensional design freedom. 
Many diverse quantum phenomena \cite{longhi2009,szameit2010} can be observed and simulated by means of such structures: in particular, a powerful analogy can be exploited between the Schr\"{o}dinger equation, describing the evolution of a wavepacket in a two-dimension potential, and the equations describing the paraxial evolution of light into a dielectric structure, such as a waveguide array. In particular, an array of coupled waveguides is equivalent to a two-dimensional array of quantum wells. The temporal evolution of a single quantum particle, placed initially in a certain well, can be mapped to the spatial evolution along the propagation direction of a single photon, injected initially in a certain waveguide. It is noted that interference is essentially a single-particle phenomenon \cite{Dirac}, even when considering coherent fields; as a consequence, this kind of experiments is performed with laser light to simplify the measurements.

Here, we investigate the role of a partial suppression of interference effects in the transport dynamics through maze-like graphs. In particular, we theoretically demonstrate that an optimal mixing of classical and quantum dynamics leads to a remarkably efficient transmission of energy/information from the input to the exit door of a generic maze. In addition, we show that it is possible to reproduce experimentally these dynamics in a photonic simulator, unfolding the maze onto a femtosecond-laser-written three-dimensional waveguide array, where noise is implemented by modulating the propagation constants of the waveguides during the writing process. The results provide a clear demonstration that a controlled amount of decoherence in the walker can produce an enhanced transport efficiency in escaping the maze and that these phenomena can be investigated in an experimentally accessible platform and not only in abstract models.

\section*{Theory}

The maze structure is created here by the so-called random Depth-First Search (DFS) algorithm applied on a square lattice of $N$ nodes \cite{shimon} (see details in S.I.). The transport model is hence represented by a walker entering the maze in some initial (IN) site or input door and moving over the structure until reaching a final (OUT) site or exit door (maze's solution). 
\begin{figure}[t]
\centerline{\includegraphics[width=.6\textwidth]{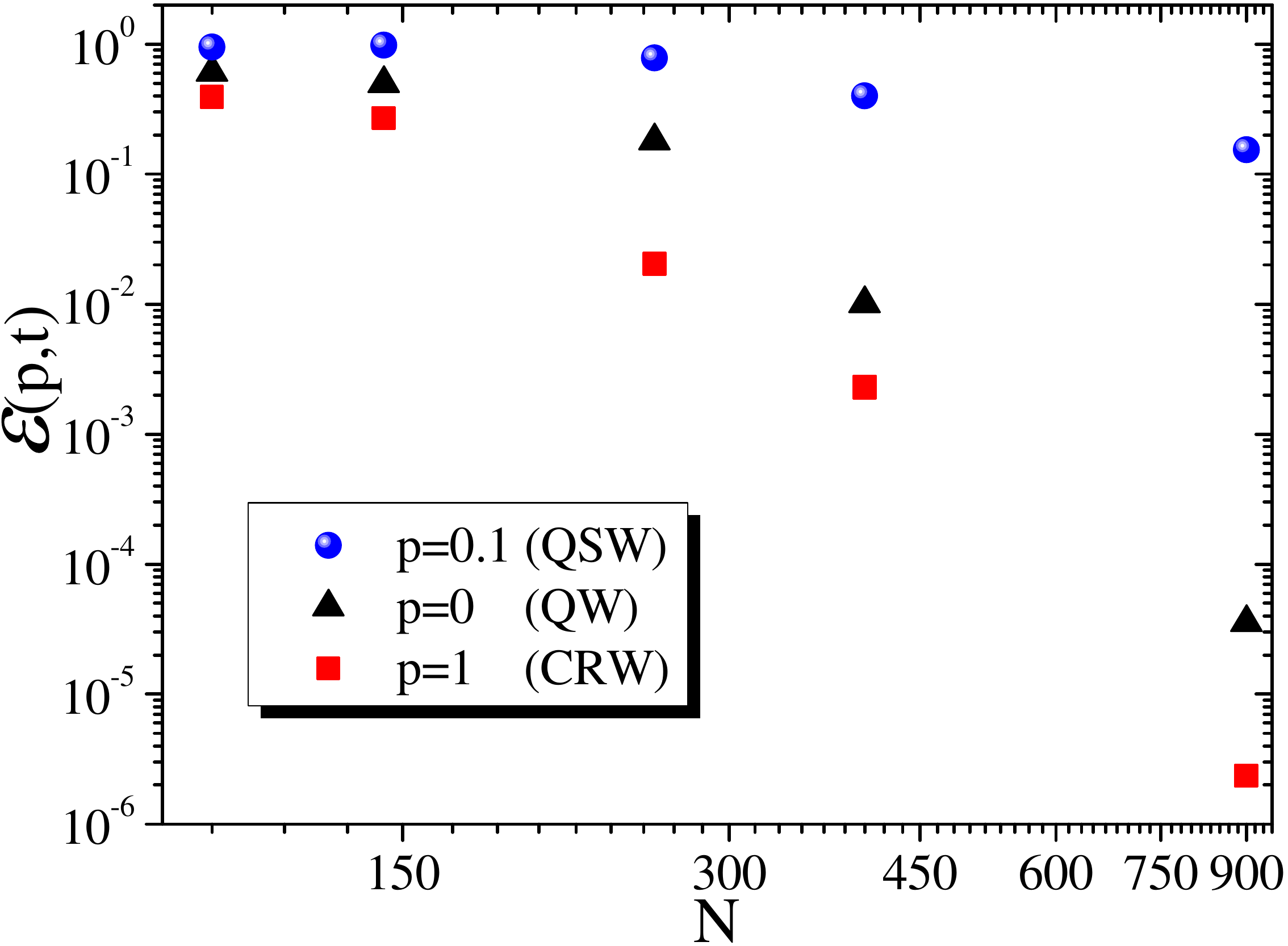}}
\caption{Transfer efficiency ${\cal E}(p,t)$ as a function of the size $N$ of the maze, for a time scale $t$ linearly increasing with $N$, i.e. $t=10 \ N$. For a maze with $N=900$ nodes, the optimal mixing $p=0.1$ provides a transfer efficiency that is about five orders of magnitude larger than the perfectly coherent (quantum, i.e. $p=0$) and fully noisy (classical, i.e. $p=1$)  regimes. The trend of the curves with the maze complexity $N$, for the different values of $p$, indicates that even higher speed-up can be achieved for increasingly larger mazes.}\label{fig2}
\end{figure}
Following the framework of quantum stochastic walks \cite{Alan2010,caruso14}, the density matrix $\rho$ describing the state of the system evolves according to the Lindblad master equation:
\begin{equation}
\label{QSRWs}
\frac{d \rho} {dt} = - (1-p) i [H, \rho] + p \sum_{i j} \left(L_{i j} \rho {L}^{\dagger}_{i j} - \frac{1}{2} \{{L}^{\dagger}_{i j} L_{i j} , \rho \} \right) \; ,
\end{equation}
A purely unitary evolution, given by the hermitian Hamiltonian $H$ which implements the quantum walk dynamics, is mixed with an incoherent evolution describing a classical random walk, given by the operators $L_{i,j}$. The balance between the two parts of the Lindblad superoperatore is given by the value of the parameter $p$. In particular, for $p=0$ a fully coherent (pure interference) dynamics is observed, while $p=1$ corresponds to the case of classical random walk, i.e. classical random hopping with no interference; for intermediate values a mixing of the two types of behaviour is obtained.
An irreversible transfer process from the exit site to an external sink is added and the walker's probability in getting the exit at time $t$ is quantified by transfer efficiency function to the sink ${\cal E}(p,t)$, whose values are in the range $[0,1]$. Further technical details are given in the S.I..

As shown in Fig. \ref{fig2}, the transfer efficiency for a quantum maze of about one thousand sites, for a given time (linearly increasing with the maze size), is more than $5$ order of magnitudes larger when one partially suppresses interference effects ($p=0.1$, i.e. about $10\%$ of mixing), with respect to the limiting cases of purely coherent and fully classical dynamics. Such transport enhancement is based on an intricate interplay between coherence and noise and shows peculiar features that makes it a fascinating field to investigate. In fact, an analogous optimal mixing has been very recently observed over a large family of complex networks for $p \sim 0.1$ \cite{caruso14}. In addition, noise-enhanced transport dynamics was observed even for totally regular and ordered graphs \cite{benjamin14} (where an intuitive picture of this optimality can be given in terms of a 'momentum rejuvenation'), thus evidencing how this phenomenon cannot be explained as just a cross-over from disorder-induced coherent localization towards classic diffusive regime.
One can consider how in the noiseless case the particle undergoes discrete diffraction in the structure: the strong interference effects given by full coherence generate bright and dark zones, even if the wavefunction does not strictly localize, and this may limit the transfer efficiency between two distant sites of the graph. Adding an optimal quantity of noise may help in suppressing the fine-grained interference pattern while keeping the wavefunction spread almost as in the ballistic case, without reaching the diffusive limit where the transport dynamics is much slower. 

\section*{Experimental realization}
\begin{figure}[ht]
\centerline{\includegraphics[width=.95\textwidth]{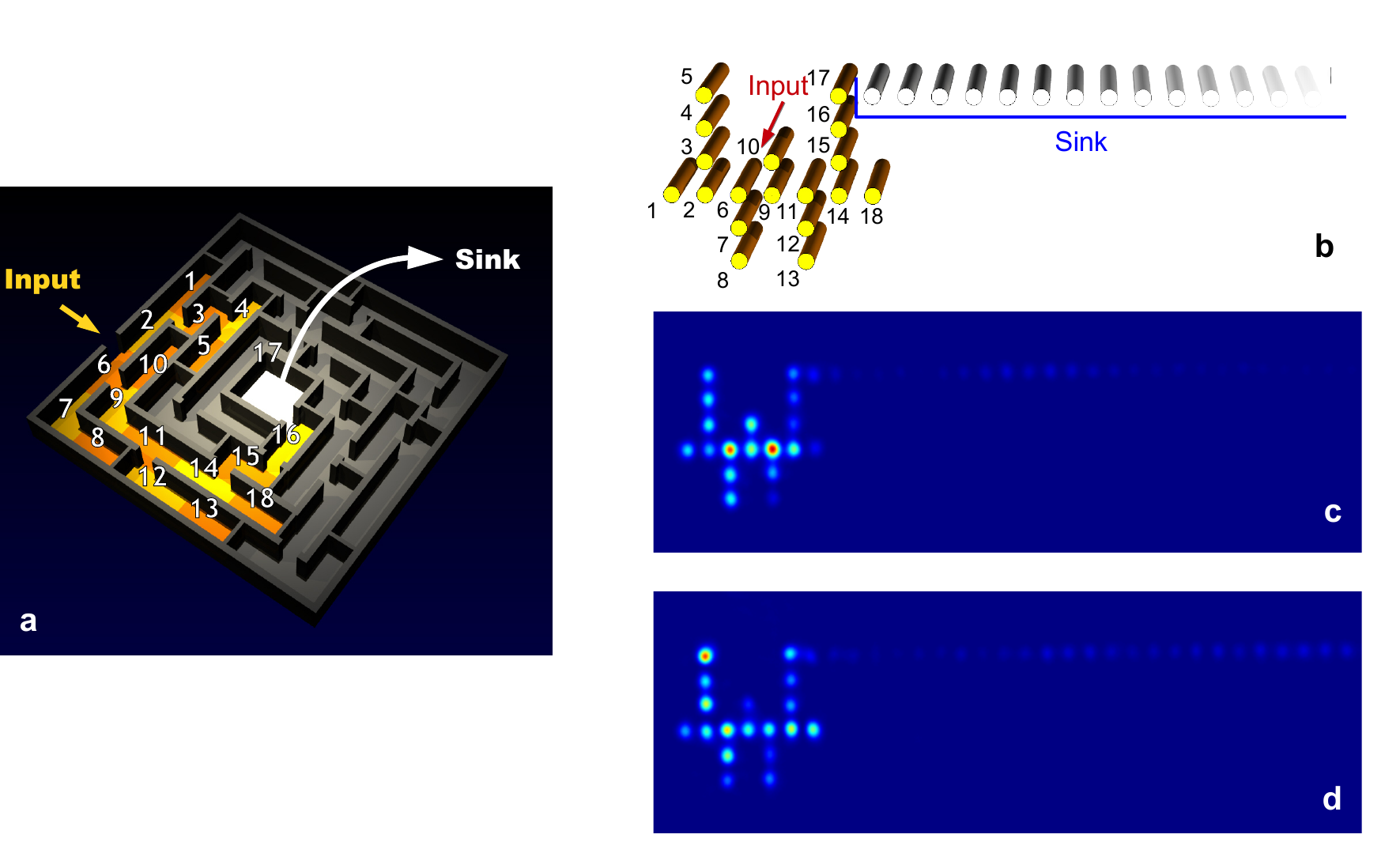}}
\caption{(a) Maze structure that is experimentally implemented; (b) unfolding of the maze into an almost linear graph, where each node is represented by a wave guide; (c-d) snapshots of the light diffusion for uniform (c) and noisy structure (d). The latter pictures correspond both to a propagation length of 60~mm. The noisy configuration is \textit{noise3} in Fig.~\ref{fig4}.}\label{fig3}
\end{figure}

Taking advantage of the unique three-dimensional fabrication capabilities of femtosecond laser waveguide writing, we implement a simulator of quantum stochastic walks by engineering an integrated photonic device probed by coherent light. The maze structure is mapped onto a three-dimensional waveguide array, in which each waveguide represents a site of the quantum maze. In particular,  our experimental study is focused on the maze configuration shown in Fig.~\ref{fig3}a, composed of 18 sites, taken as a significant example for observing the dynamics predicted by our theoretical model (see S.I. for experimental details).

A first problem that has to be addressed is how to map in a waveguide system the topology of the links between the sites of our maze. Whereas in an arbitrary maze structure transfer between adjacent sites can be inhibited by walls, in waveguide arrays the coupling between two waveguides is solely determined by their relative distance. Thus, the geometry of the array needs to be engineered to keep far enough from each other waveguides that must not couple. This might not be possible if the maze graph is too complex.
In our case, however, it was possible to unfold the maze graph in Fig.~\ref{fig3}a, by considering chains with side tails,
onto the partially linear and more feasible structure in Fig. \ref{fig3}b. Note that this unfolded geometry is not unique, other configurations being conceivable in principle with the same distances between equally coupled sites.

Another experimental issue is the realization of the exit door (i.e. OUT site). In the theoretical model this site should behave like a sink that absorbs energy irreversibly. In our photonic implementation the sink is implemented by a long chain of waveguides (62 waveguides) which well approximate a one-way energy transfer process, with negligible probability for the light to be coupled back to the system.

Structures composed of uniform waveguides correspond to the purely coherent case (QW). Fully coherent transport dynamics in such maze can be studied straightforwardly by fabricating arrays with different lengths and characterizing the output distribution when coherent light is injected in the desired initial site (IN). It is worth noting that in this realization, the evolution parameter $t$, considered in the theoretical model, is mapped onto the propagation length, which we still label as $t$.

A controlled amount of noise is introduced in the structure by segmenting the waveguides corresponding to the sites of the maze. This is achieved by modulating the writing speed in the fabrication process, which induce a proportional variation of the propagation constant, while keeping the coupling coefficient unvaried \cite{CorrNC}. The value of the propagation constant variation in each segment is randomly picked from a uniform distribution with a given amplitude; the same distribution is used for every waveguide within the same array. The random variation of the propagation constants is equivalent to a random variation of the site energy due to the interaction with an incoherent environment \cite{MRLA2008}. This approach has been extensively tested by numerical simulations of this specific implementation as compared to the theoretical Lindblad model discussed in the previous section (see the discussion in the SI for details).

The waveguide array implementing the sink is in all cases composed by uniform, not-segmented, waveguides. To characterize the transfer efficiency to the sink, the output facet of each fabricated structure is imaged onto a CMOS camera (examples of snapshots are shown in Figs.~\ref{fig3}(c-d) ), the light intensity on the maze and sink regions of the array are numerically integrated and the fraction of light in the sink is calculated.

Twenty-four structures were fabricated with the transverse layout as in Fig.~\ref{fig3}b, implementing six different propagation lengths for both the noiseless, fully-coherent, situation and three different noise configurations with the same strength (i.e. same amplitude of propagation constant distribution). The coupling coefficient between nearest-neighbouring waveguides is $\kappa = 0.40$~mm$^{-1}$. The amplitude of the random distribution of the propagation constants, adopted in the noise implementation, is $\Delta \beta_{max} = 0.40 $mm$^{-1}$. The value of the propagation constant is modified every 3~mm of waveguide length.

Note that the length of the segments with fixed propagation constant may be related to a sort of coherence time of the walker, i.e. an evolution time in which the phase relations are conserved. The overall dynamics depends non-trivially on the interplay of this length and the amplitude of the random distribution, together with the characteristic interaction length of the system, given by the coupling constant. From a practical point of view, the 3-mm value was chosen as the minimum allowed by the acceleration capability of our translation stages.

\begin{figure}[t]
\centerline{\includegraphics[width=.85\textwidth]{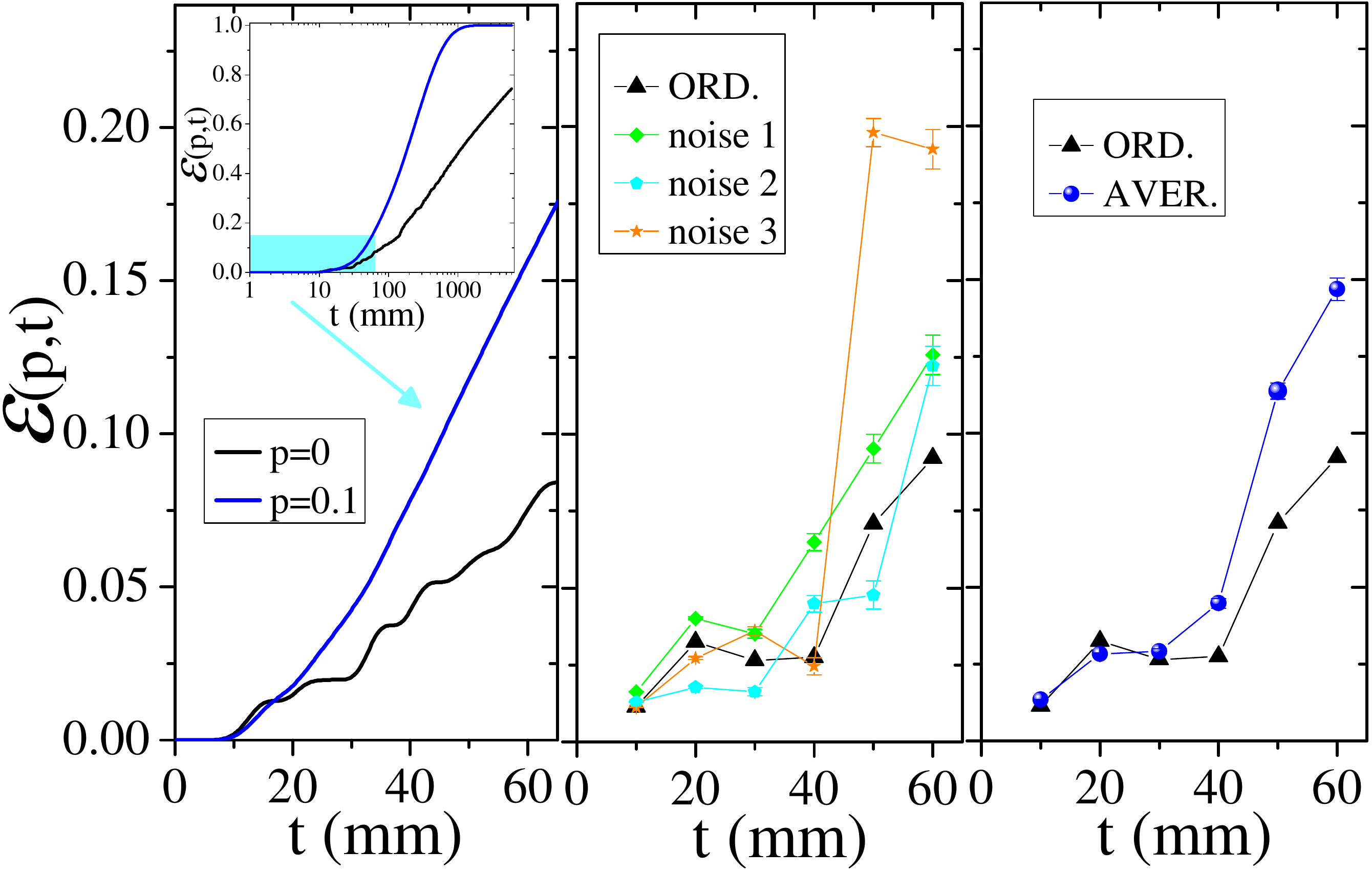}}
\caption{Left: Theoretical behaviour of the transfer efficiency ${\cal E}(p,t)$ as a function of the evolution parameter $t$ for two values of the mixing parameter $p$, corresponding to QW ($p=0$) and QSW ($p=0.1$), for the maze in Fig. \ref{fig3}. 
Inset: a larger time scale is shown in order to point out the remarkable efficiency enhancement when there is a partial suppression of interference. 
Center: Experimental results for both ordered (triangles) and three noisy realizations of the structure reported in Fig. \ref{fig3}. Where not shown the error bars are smaller than the symbol. Right: as in the middle panel, but only with the ordered case (triangles) and with the average efficiency over the noise realizations (circles).
In all three panels, the time $t$ is in units of $mm$ because it is experimentally mapped into the propagation length of the three-dimensional waveguide array.}\label{fig4}
\end{figure}

It is worth noting that in this realization, the evolution parameter $t$, considered in the theoretical model, is mapped onto the propagation length, which we still label as $t$.

\section*{Results and discussion}

The transfer efficiency to the sink, calculated theoretically with the method reported in \cite{caruso14}, is shown in Figure~\ref{fig4}a for a maze with the layout presented in Fig.~\ref{fig3}. Such layout represents the actual structure that we have experimentally implemented and is considered both for the case of fully coherent quantum transport and for the case of partially incoherent transport with $p=0.1$. Figure~\ref{fig4} shows the experimentally retrieved transfer efficiencies, each point corresponding to a physically different structure fabricated to implement a certain noise map and a given propagation length. The average between the points corresponding to the three different noise implementations is also shown (right panel).

The physical quantity that is measured experimentally is the fraction of light \textit{present} in the sink after a certain propagation, and not the fraction of light that is \textit{transferred} to the sink. In case propagation losses are the same both in the maze waveguides and in the sink waveguides, the two quantities indeed correspond. As a matter of fact, the modulation of the writing speed produces additional losses in the waveguides of the maze with respect to the waveguides of the sink and this causes in general an overestimation of the transfer efficiency. However, we characterized accurately such additional losses in our structures and simulated their impact on the estimation of the transfer efficiency (see S.I.). The consequent systematic error contribution has been directly taken into account in Fig.~\ref{fig4}, while the random error contribution is reported with the error bars.

A very good agreement between theoretical and experimental curves is observed both for the noiseless, fully coherent case, and for the partially coherent transport when considering the average of our ``noisy'' waveguide implementations. Therefore, this experimental evidence validates our claim that the interplay of noise and interference effects leads to higher efficiency in finding the maze solution. 

To summarize, here we have studied both theoretically and experimentally the dynamics of a walker traveling in a maze, having a single path from the input door (starting point) to the exit (solution). By considering a model that mixes the behavior of a classical walker and a quantum one, we have found an optimal condition leading to extremely efficient and fast transmission. For large enough maze size this leads to a remarkably high enhancement of more than five order of magnitudes in the transfer efficiency with respect to both the classical and purely quantum limits. 

By exploiting the unique capabilities of the femtosecond laser writing technology we have unfolded the maze and implemented it in a three-dimensional waveguide array, where a suitable modulation of the waveguide properties allowed us to mimic a partial decoherence of the walker. Our measurements have faithfully confirmed the theoretical predictions and, in particular, the remarkable role of a partial suppression of interference in enhancing transport dynamics in mazes. It is also worth noting that our technological platform has enabled an experimental implementation of a noise-assisted problem in well-controlled conditions and over complex topologies, and can thus represent a very powerful tool for further studies in this direction. These results pave the way to much more powerful integrated photonics devices exploiting interference, quantum features and noise effects for improved problem solving efficiency, and remarkably fast transmission of information in ICT applications and of energy in novel solar technologies.

\section*{Authors contribution}
F.C., F.S. and R.O. conceived the whole project. A.C., A.G.C. and R.O. carried out the experiment; F.C. led and carried out the theoretical work. A.C. and A.G.C. exploited ultra-fast laser writing process to realize the maze structure. F.C. and A.C. performed the numerical simulations of the transport dynamics. F.C., F.S. and R.O. supervised the project. All authors contributed to the discussion, analysis of the results and the writing of the manuscript.

\section*{Acknowledgments}
We acknowledge M.B. Plenio for carefully reading the manuscript and very useful and stimulating discussions.
This work was supported by the European Union through the project FP7-ICT-2011-9-600838 (QWAD —Quantum Waveguides Application and Development), European Research Council (ERC-Starting Grant 3D-QUEST, 3D-Quantum Integrated Optical Simulation, grant agreement no. 307783, http://www.3dquest.eu/), EU FP7 Marie-Curie Programme (Career Integration Grant, project no. 293449), and by national grants as PRIN (Programmi di ricerca di rilevante interesse nazionale) project AQUASIM (Advanced Quantum Simulation and Metrology) and MIUR-FIRB project (RBFR10M3SB). We acknowledge QSTAR for computational resources, partially based also on GPU programming by NVIDIA Tesla C2075 GPU computing processors.

\section*{Supplementary Information}

\subsection*{Theoretical model for quantum stochastic walks}

The analysis of the transport dynamics on the maze structure is theoretically supported by the the general framework of quantum stochastic walks \cite{Alan2010,caruso14}, where the physical state of the system is described by the density matrix $\rho$ and follows this evolution:
\begin{equation}
\label{QSRWs}
\frac{d \rho} {dt} = - (1-p) i [H, \rho] + p \sum_{i j} \left(L_{i j} \rho {L}^{\dagger}_{i j} - \frac{1}{2} \{{L}^{\dagger}_{i j} L_{i j} , \rho \} \right) \; ,
\end{equation}
where $H$ is the Hamiltonian including the hopping parameters $T_{ij}$ between the connected nodes $i$ and $j$ described by the states $| i \rangle$ and $| j \rangle$ (all having the same energy), while the operators $L_{i j} = T_{ij} | i \rangle \langle j |$ are responsible for the presence of the effective dephasing noise experimentally induced by dynamical disorder on the integrated structure. The limit $p=0$ corresponds to the case of 'no noise", while $p=1$ is obtained when the interference effects are completely destroyed and the hopping is just classical. For other values of $p$ between $0$ and $1$ there is a mixing of the two types of dynamics, i.e. combining classical and quantum effects.

\begin{figure}[t]
\centerline{\includegraphics[width=12cm]{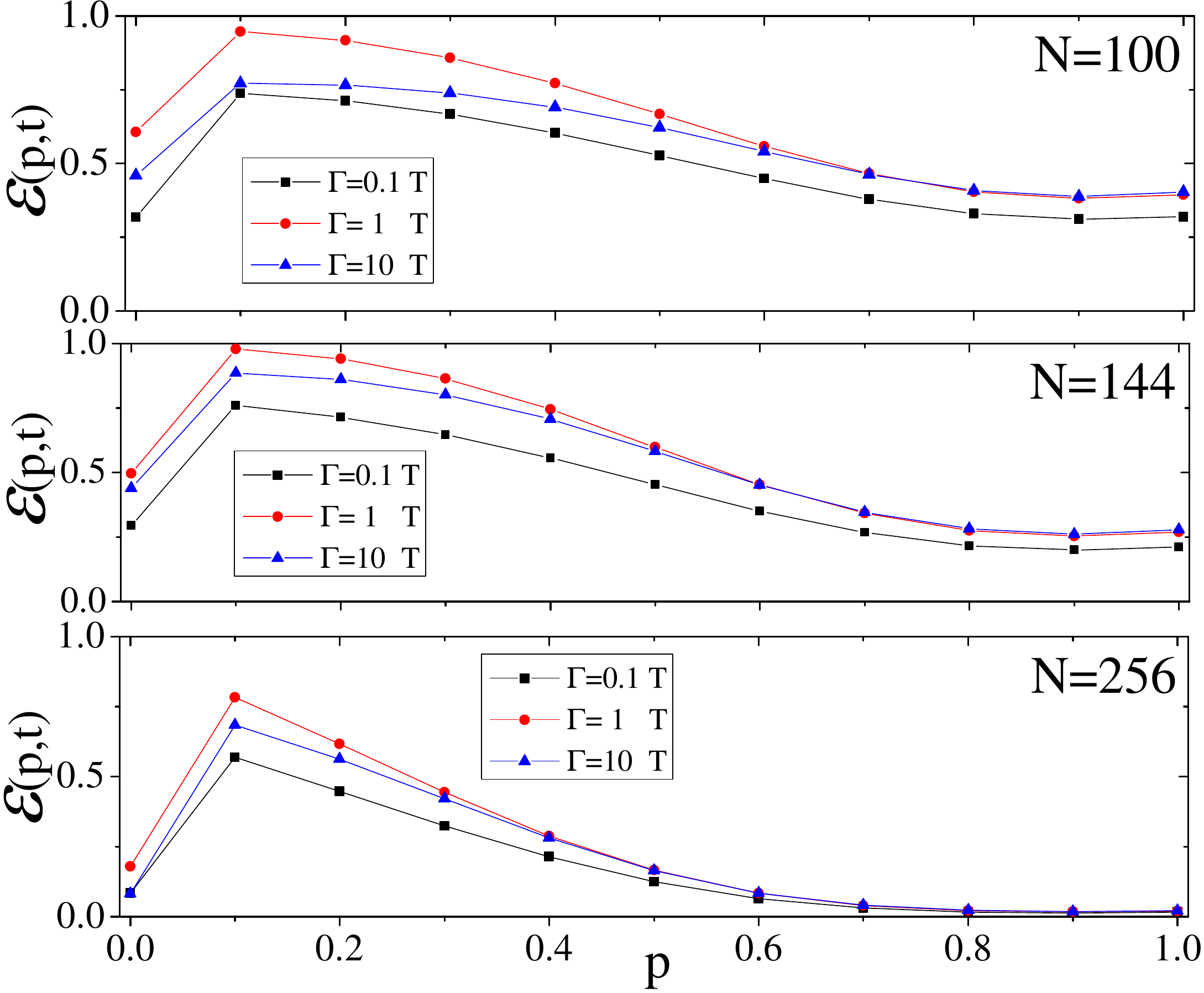}}
\caption{Transfer efficiency ${\cal E}(p,t)$ as a function of $p$, for a time scale $t$ linearly increasing with $N$, i.e. $t=10 \ N$, for mazes of size $N=100$, $N=144$ and $N=256$ (from top to bottom). In each panel the transfer rate $\Gamma$ is varied by two order of magnitudes in order to show the robustness of our results with respect to the specific choice of $\Gamma$.}\label{SI-FIG1}
\end{figure}

The transfer efficiency of the structure is measured by adding, on the right-side of Eq. (\ref{QSRWs}), another term $L_{N+1} = \Gamma \ [ \ 2 |N+1 \rangle \langle N| \ \rho \ |N \rangle \langle N+1| - \{|N \rangle \langle N|,\ \rho\} ]$, with $\Gamma$ being the irreversible transfer rate from the site $N$ of the graph into some external node $N+1$ (exit door or OUT site) where the energy is continuously and irreversibly stored. Hence, the transfer efficiency of the quantum maze will be measured by \cite{CCDHP2009,caruso14}
\begin{equation}
{\cal E}(p,t) = 2 \Gamma \int_{0}^t\rho_{NN}(p,t')\mathrm{d}t' \; .
\end{equation}
where $\rho_{NN}(p,t')$ is the population of the site $N$ at time $t'$ in the case of a mixing strength $p$.
In this manuscript the model parameters are:  1) uniform nearest-neighbor couplings, i.e. $T_{ij}  \equiv T$ for any connected nodes $i$ and $j$, 2) next-nearest neighbor couplings between the sites in the main chain and the closest ones on the relative tails $T_{ij}  \equiv 0.2 \ T$ (these are sites placed at 45$^\circ$, at a relative distance $\sqrt{2}$ times larger than the first neighbors), to match with the experimental observations, 3) $\Gamma=T$, 4) the time is measured in units of $T^{-1}$. 
Let us point out that the optimality of $p \sim 0.1$ does not depend on the specific choice of $\Gamma$. Indeed, by changing $\Gamma$ by two orders of magnitudes, i.e. in the range $[0.1, 10] \ T$, the optimal transport efficiency is still at $p \sim 0.1$ -- see Fig. \ref{SI-FIG1}.
We note that in the model above we have a mixing of classical hopping process and quantum walks, leading to the suppression of interference effects, that is instead experimentally implemented by means of a qualitatively different process, i.e. introducing dynamical disorder in the waveguide structure along the propagation length. However, this does not affect our results, because, by numerically considering the presence of dynamical disorder (instead of the mixing $p$ component), we find a very similar transport efficiency behaviour -- more details in the section below on the Photonic model and in Fig. \ref{SI-FIG3}.

\subsection*{Maze construction}

\begin{figure}[t]
\centerline{\includegraphics[width=1.\textwidth]{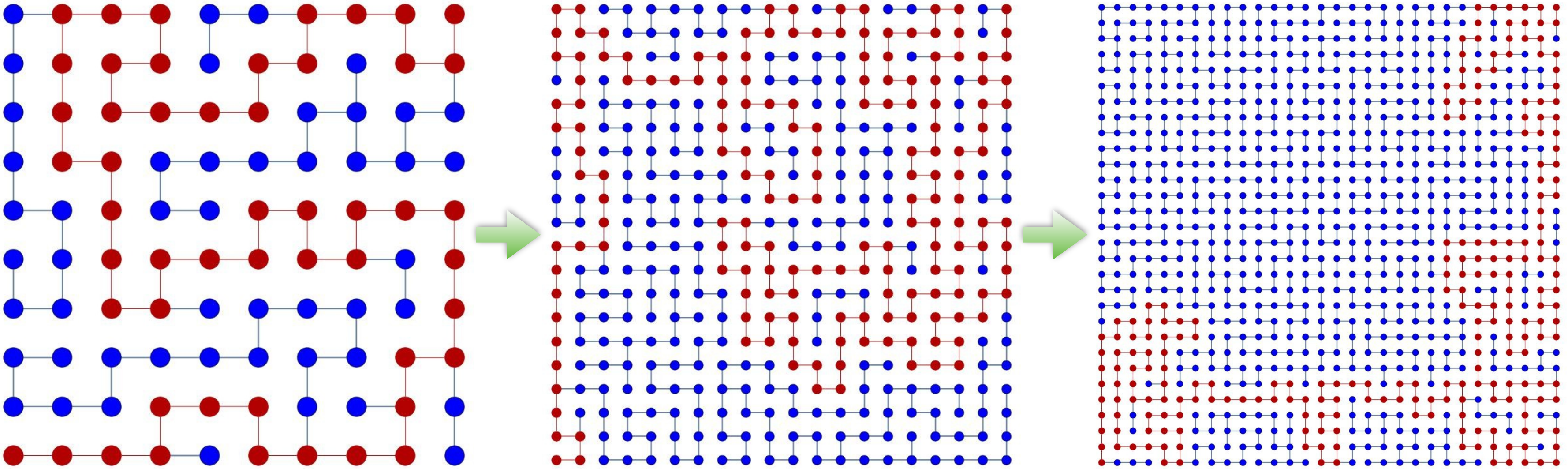}}
\caption{Making the maze larger and larger by applying DFS algorithm to a square lattice of higher number $N$ of nodes. The only pathway connecting the IN and OUT sites (placed at the opposite corners of the lattice) is shown in red.}\label{SI-FIG2}
\end{figure}

Depth-First Search  (DFS) algorithm is the simplest maze generation algorithm and is based on the following iterative procedure that is applied to a regular square grid of $N$ nodes, where all neighbor sites are separated by a wall \cite{shimon}. One starts from a random node and then search for a random neighbor that has not considered yet. If so, the wall between these two sites is knocked down; otherwise, one backs up to the previous node. This procedure is repeated until all sites of the grid have been visited. By doing so, the final structure is a maze where we have only one path connecting the IN to the OUT node, i.e. a so-called two dimensional perfect maze with no closed loops. Applying this procedure to larger and larger square lattices, one obtains increasing large maze graphs -- see Fig. \ref{SI-FIG2}.

\subsection*{Photonic model}

Before fabricating the waveguide devices, we investigated accurately by numerical simulations the possibility to implement quantum walks with mixtures of quantum and classical behaviours by our photonic structures. 

Mazes with different sizes and topologies were considered. First they were unfolded as described in the Main Text, to give the geometry of feasible photonic structures. A long waveguide chain was added, coupled to the output site, to implement the sink. The propagation of coherent light, injected initially in a chosen input site, was simulated by coupled mode equations of the kind \cite{perets}:
\begin{equation}
i \frac{d A_n}{d z} = \Delta \beta (z) A_n + \sum_{m \neq n} \kappa_{m,n} A_m
\end{equation}
where $A_n$ is the field amplitude on the $n$-th mode, $\Delta \beta (z)$ is the detuning of the propagation constant of the waveguide, which varies randomly in segments along z to implement noise, $\kappa_{m,n}$ is the coupling coefficient between the $m$-th and the $n$-th waveguides. Coupling coefficients between third-neighbour waveguides and above were considered negligible. 
The transfer efficiency of the quantum maze was calculated as the fraction of power being in the sink waveguide array.

\begin{figure}
\centerline{\includegraphics[width=1.\textwidth]{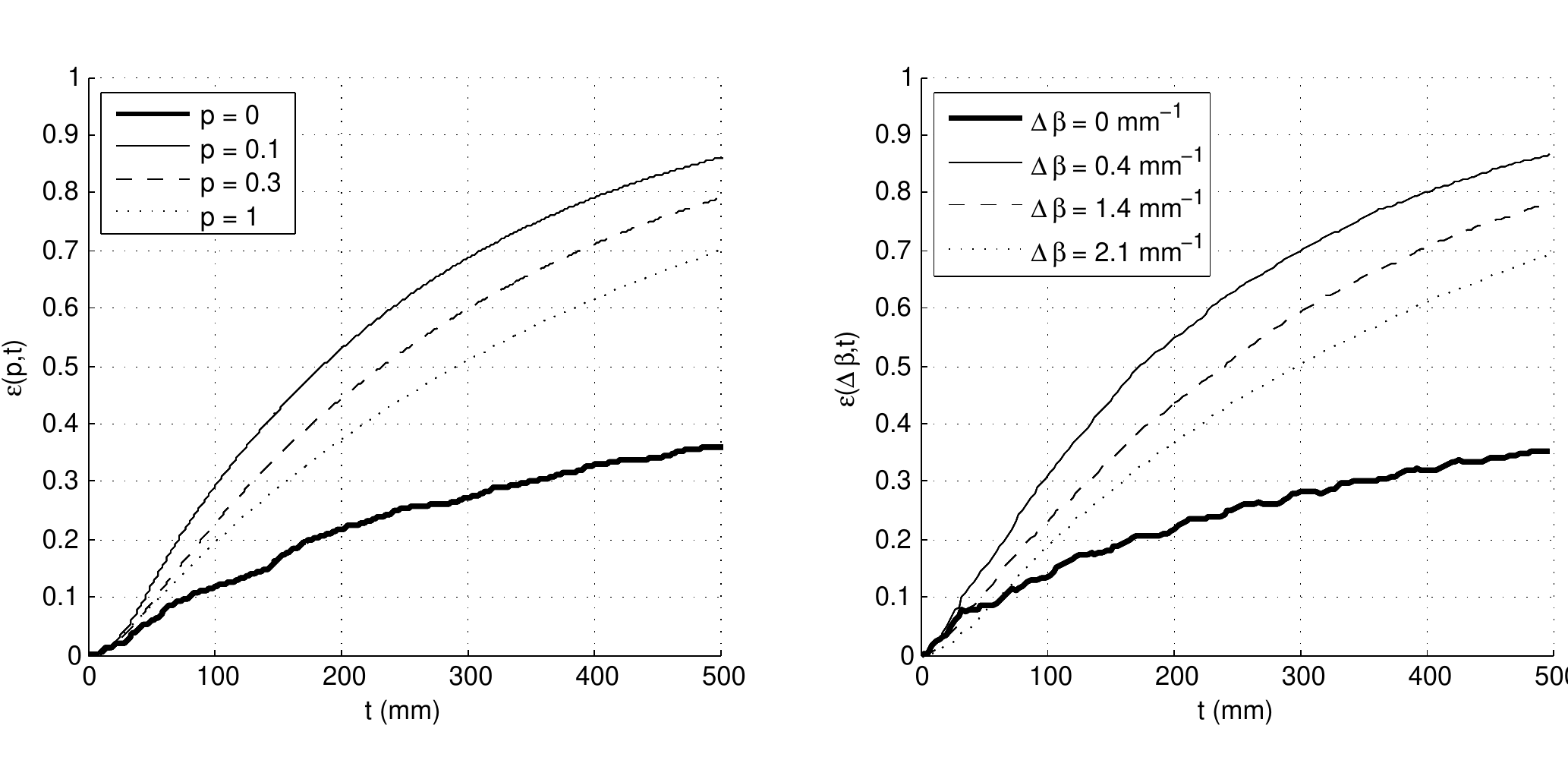}}
\caption{Transfer efficiency as a function of the propagation length $t$, calculated with a quantum stochastic walk model for different values of $p$ (left), and with a photonic simulation for different amplitudes of the random distribution of $\Delta \beta$ (right). 
For the photonic simulation the average over 100 random distributions of $\Delta \beta$ with the same amplitude is shown.}
\label{SI-FIG3}
\end{figure}

The results of this kind of numeric simulation were compared with the behaviour predicted by the quantum stochastic walk model described above.   An excellent agreement between the two procedures was evidenced: the behaviour obtained for a certain value of the parameter $p$ of the quantum stochastic walk can be faithfully reproduced with a proper amplitude of the random distribution of $\Delta \beta$ in the photonic model. As an example, Fig.~\ref{SI-FIG3} reports the transfer efficiency obtained with a quantum stochastic walk model and with the coupled-mode equations described above, for a structure with the topology of Fig.~3 of the Main Text.

\subsection*{Mimicking decoherence}

The Lindblad model discussed in the first section and described by Eq.~\ref{QSRWs} is a model which generates an interplay between a  quantum walk and a classical random walk with a proportion given by the phenomenological parameter $p$. While the quantum-walk part of the Lindblad superoperator connects the \textit{population} (diagonal) terms of the density matrix with the \textit{coherence} (off-diagonal) terms, the classical part connects directly the population terms one to the other. Given an initial pure quantum state, if only the quantum-walk part were present, it would remain a pure state through the propagation; in contrast, the classical part makes it a partially mixed state.

It should be noted that the classical part in the Eq.~\ref{QSRWs} does not refer to a particular microscopic process: in contrast, in a real physical system the interaction with the environment will have its peculiar microscopic characteristics. In our realization, we aim at implementing an interaction with the environment that effectively varies the energy of the sites composing the maze, with random variations around a "zero" level (similarly also to Ref. \cite{MRLA2008}). Since the environment can be seen as an incoherent system, such random variations are purely \textit{classical}, modelled with actual variations of the energy of the sites in specific noise realizations.
Thus, as discussed in the Main Text and in the previous section, a controlled amount of decoherence in the walk is mimicked by noise in the propagation constant; i.e., a waveguide is composed of adjacent segments of fixed length $L$ with different propagation constants (uniform inside each segment) with values randomly picked from a uniform distribution with a given amplitude $\Delta \beta_{max}$. 

Note that at the end of each segment and at the end of the propagation, the specific noise map implemented (which is random in a classical sense) determines the pure state that we measure in the end. The population terms of such pure state are characterised with our measurements with classical light. To reconstruct the full mixed state a rich statistics of noise realizations would be needed. However agreement with the Lindblad model (which does not refer to single realization but general statistical properties of the ensemble) can be checked with a small number of realizations.

\begin{figure} [t]
\centerline{\includegraphics[width=1\textwidth]{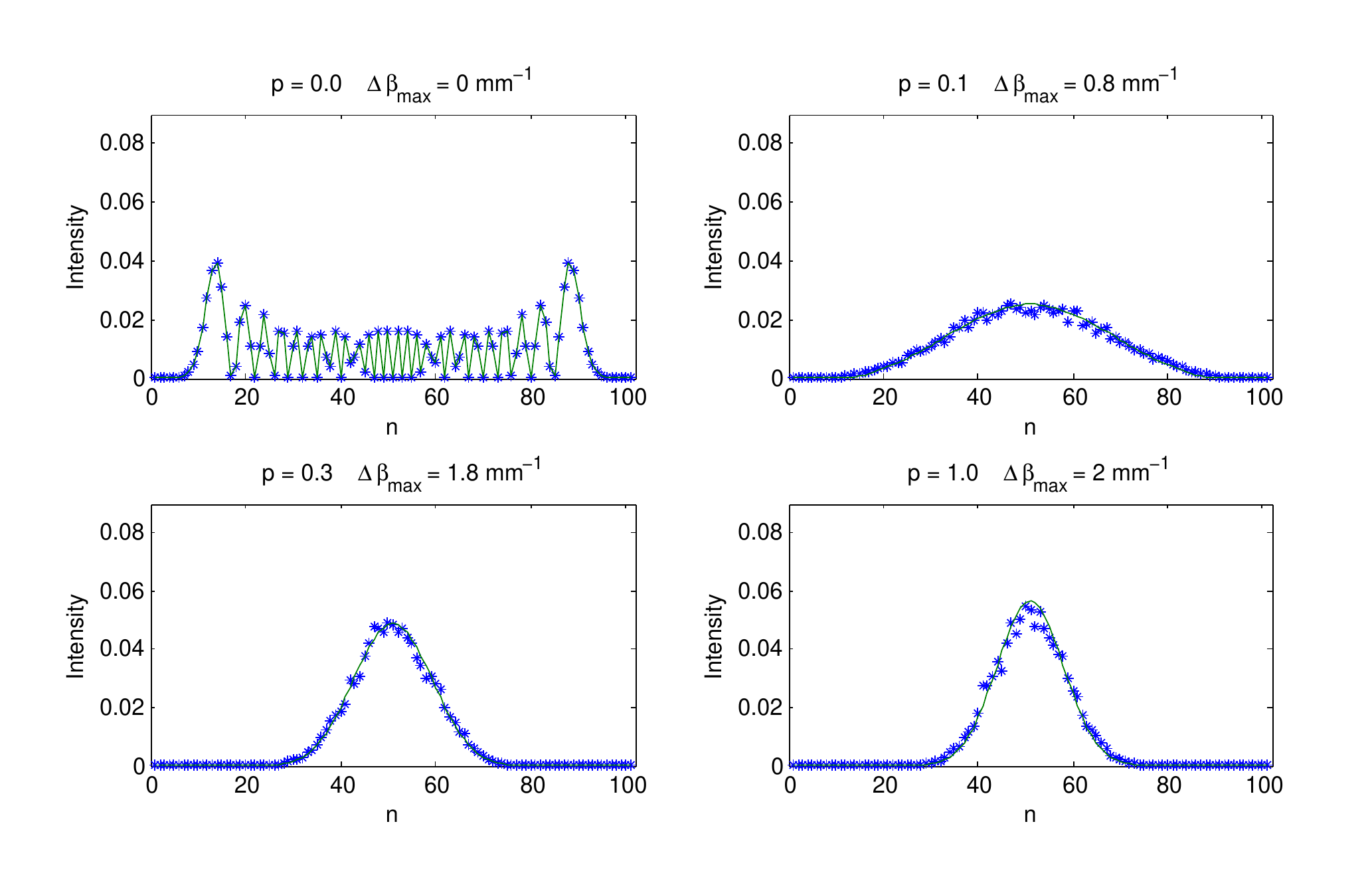}}
\caption{(stars) Simulated averaged output distribution for linear waveguide arrays of 101 waveguides, after 50~mm propagation. Propagation constants of the waveguides are modulated randomly in segments of 3~mm length, picking the values from uniform distributions of amplitude $\Delta \beta$. The coupling coefficient between adjacent waveguides is 0.4~mm$^{-1}$. The average is performed on 200 noise realizations. (continuous line) Simulation with the Lindblad model for an equivalent coupling between the sites and different values of $p$.}
\label{SI_FigSQRW}
\end{figure}

The possibility to use the Lindblad formalism for describing this kind of microscopic dynamics implemented in detail by a waveguide array  with random variations of the propagation constants, or, conversely, the possibility to use such a kind of photonic system to experimentally investigate the dynamics of the Lindblad model, is further confirmed from the data in Fig. \ref{SI_FigSQRW}. The figure shows the simulated output distributions (averaged on 200 noise realizations) after 50~mm propagation in linear arrays of 101 waveguides, where the propagation constants are modulated randomly as described above and coherent light is injected in the central waveguide, compared to simulations for different values of $p$ for the Lindblad model. Increasing the modulation amplitude the distribution progressively change from the typical ballistic spread of the quantum walk to the gaussian shape of the classical random walk, with very good agreement between the two models. 

It is interesting to note that even in such a completely ordered system, with no localized eigenstates in absence of noise, when analyzing the transport dynamics from the center of the array to one end, noise-assisted transport was recently observed for similar optimal noise conditions $p = 0.1$ \cite{caruso14}. As shown in Fig. \ref{SI_FigSQRW} the distribution for $p=0.1$ yields marked differences with respect to both the classical ($p=1$) and quantum ($p=0$) limiting cases, which further emphasizes the peculiarity of this phenomenon.

\subsection*{Experimental details}
Waveguide arrays are inscribed in EAGLE2000 glass substrates (Corning), by femtosecond laser writing. A Yb-based amplified laser system (FemtoREGEN, HighQLaser) provides laser pulses with 400 fs duration, 300 nJ energy and 1 MHz repetition rate, focused in the substrate by a 0.45 NA 20$\times$ microscope objective. Such objective is compensated for spherical aberrations at 170 $\mu$m below the glass surface, which is the average depth of the fabricated structures. The translation speed was varied in the 10-40 mm/s range to achieve the desired $\Delta \beta = 0.4~\mathrm{mm}^{-1}$ modulation interval. The waveguides yield single-mode operation at the wavelength of 850 nm.

The fabricated structures are probed by coherent light. Laser light at 850~nm wavelength is injected into the input waveguide. The output distribution is imaged onto a CMOS camera by means of a 0.12 NA objective.
Numerical integration on different parts of the acquired image allows to retrieve the fraction of light present in the sink waveguide array. The advantages of this method are, on one hand, insensitivity to coupling losses of the input beam and, on the other hand, the possibility of a fast acquisition of the output of many waveguides.

\begin{figure}[t]
\centerline{\includegraphics[width=.55\textwidth]{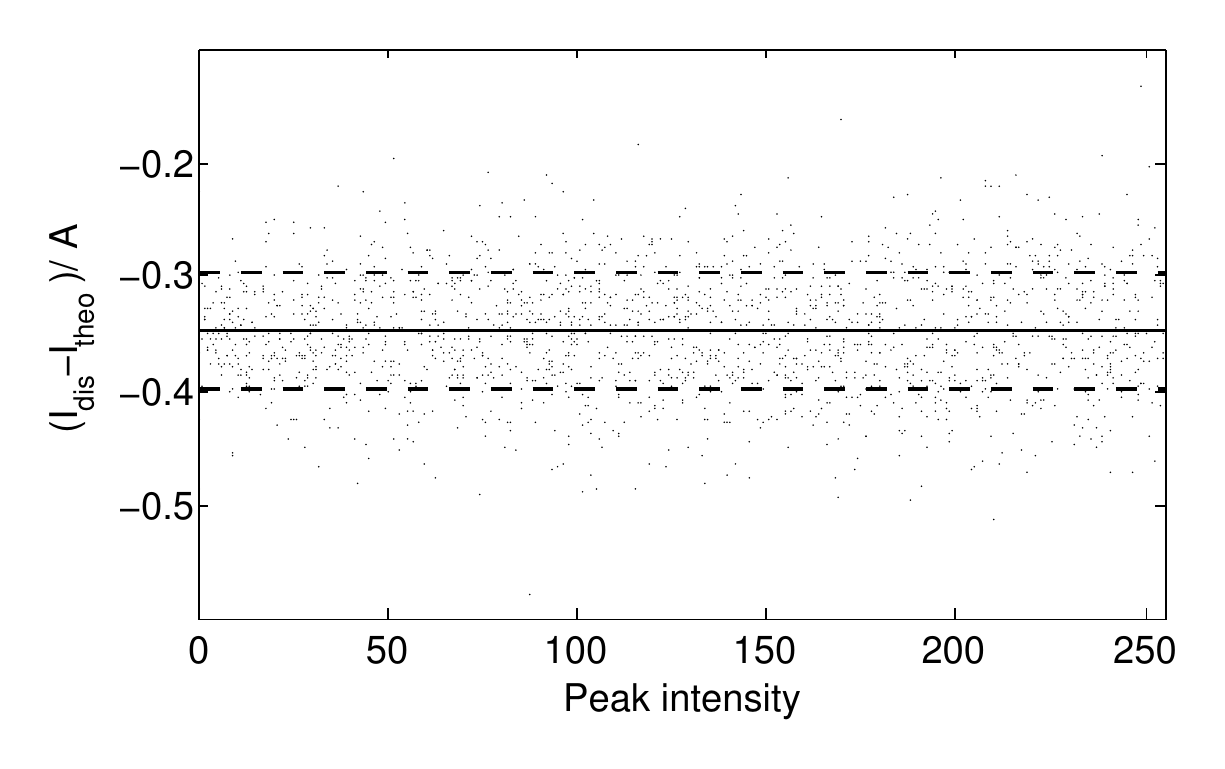}}
\caption{Difference between numerically calculated integral $I_{dis}$, affected by pixel and level discretization, and the analytical integral $I_{theo}$ of 1000 gaussian modes with random intensity and position, normalized to the mode area $A = \pi w_x  w_y$ ($w_x$ and $w_y$ are the 1/e mode radii in the two dimensions). The continuous horizontal line marks the average value ($\sim$-0.347). The two dashed lines indicate the standard deviation levels ($\sim-0.347\pm0.052$). Average and standard deviation values have been computed on $10^5$ points. }\label{SI-FIG4}
\end{figure}

A careful analysis of the measurement error has been performed. A possible source of error is the quantization of the intensity levels of the CMOS sensor, as well as its finite spatial resolution. To analyse this error contribution we simulated the numerical integration of gaussian modes, with the same size as the measured ones, but random intensity and peak position, discretized both in the (256) intensity levels and in the pixels of the spatial profile. Figure \ref{SI-FIG4} shows the (normalized) difference between the numerically calculated integral and the analytic integral of the gaussian profile, for 1000 randomly distributed modes, as a function of the peak intensity. Note that, given a certain peak intensity, the numerical integral may give different values depending on the position of the peak, because of pixel discretization. Systematic and random errors are almost independent on the peak intensity and they have been taken into account in data elaboration assuming that the acquired image contains $n$ modes of random uniformly-distributed intensity. As a matter of fact, the contribution of such errors on the experimentally measured efficiencies is relevant only for the shortest lengths, where the light intensity in the sink is particularly low.

Furthermore, as mentioned in the Main Text, cascading waveguide segments with different fabrication speed (to mimic noise) causes small additional losses, at each interface between different waveguide segments. Importantly, these losses are present only in the waveguides representing the maze and not in the sink waveguides. Because of those losses, measuring the fraction of power \textit{present} in the sink after a certain propagation distance, with respect to the overall output power, as is done in our characterization process, may not correspond precisely to the fraction of input power \textit{transferred} to the sink. In fact, the transfer efficiency is slightly overestimated.

Overall additional losses can be measured quite accurately, by simply measuring and comparing the insertion losses of the fabricated devices (ration of the overall output power over the input power). However, the precise contribution of these losses on each measured transfer efficiency can be hardly retrieved. In fact, the transfer process is not uniform during the propagation and depends on the \textit{random} noise map implemented.

\begin{figure}[b]
\centerline{\includegraphics[width=10cm]{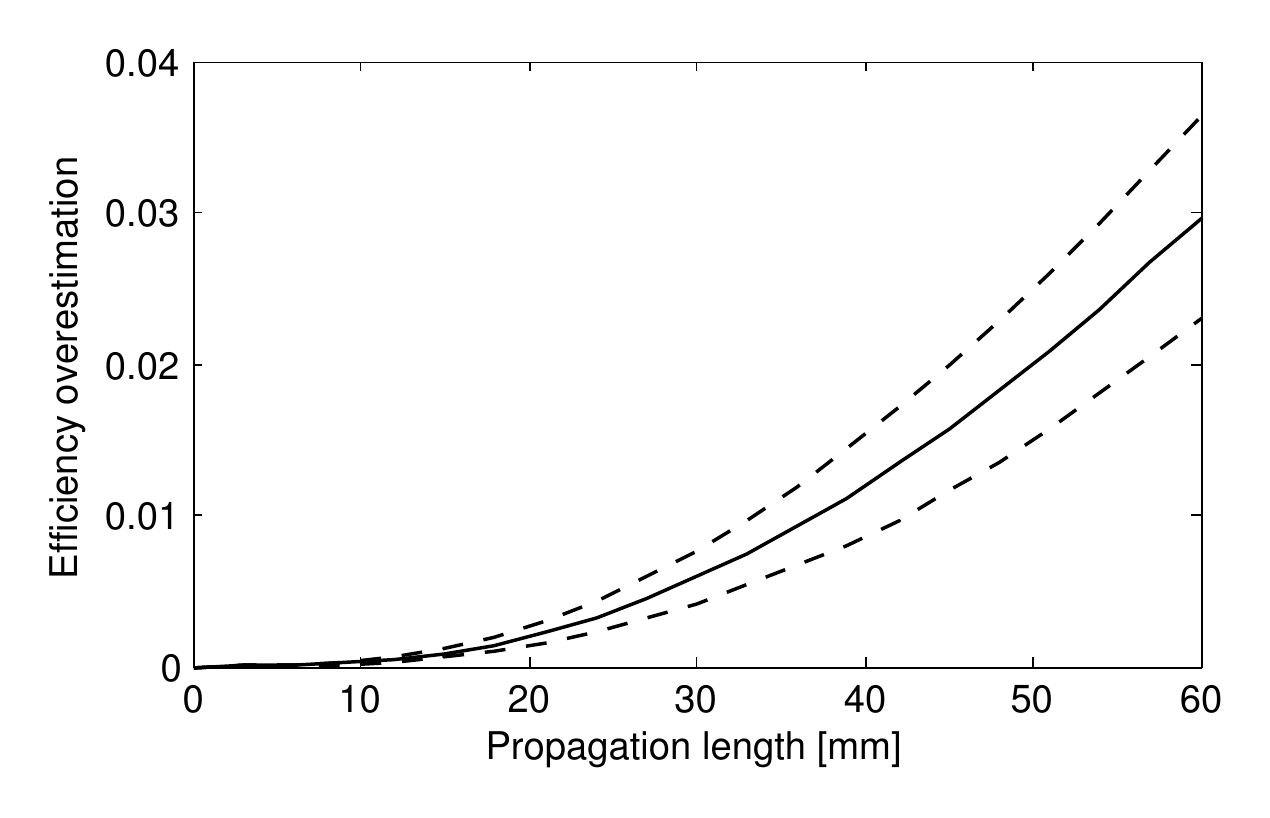}}
\caption{Overestimation of the transfer efficiency, when measured as the ration between the light present in the sink over the total output light, as a function of the propagation length. Uniform additional losses in the maze waveguides are assumed and simulations are performed on 100 random noise distributions, with amplitude $\Delta \beta = 0.4~\mathrm{mm}^{-1}$. The continuous line shows the average overestimation. The dashed lines indicate the standard deviation.}\label{SI-FIG5}
\end{figure}

Thus, to statistically quantify such overestimation we numerically simulated the light propagation in waveguides structures analogous to the fabricated ones, for 100 random different noise distributions (with always the same amplitude $\Delta \beta = 0.4~\mathrm{mm}^{-1}$, as in our experiments), both in the case of waveguides with no losses (which corresponds to the ideal situation) and in the case of waveguides yielding uniform additional losses with respect to the waveguides of the sink, in such a way that the overall losses of the structure correspond to the experimentally measured ones ($\sim$2~dB additional losses for the longest arrays). We evaluated in each case the estimation error of the transfer efficiency and calculated the error statistical distribution. 
Figure \ref{SI-FIG5} shows the average error, together with its standard deviation, as a function of the propagation length. The effect of these losses reveals to be small (the systematic component is less than 3\% for 60~mm, correspondent to our longest fabricated devices) and does not significantly influence our experimental observation of an increase in transfer efficiency in the cases in which noise is added.

\end{document}